\documentclass[aps, prl, twocolumn, superscriptaddress, floatfix]{revtex4-2}
\usepackage{graphicx}
\usepackage{amsmath}
\usepackage{ulem}
\usepackage{amssymb}
\usepackage{tikz}
\usepackage{braket}
\usepackage{float}
\usepackage{pgfplots}
\usepackage{mathrsfs}
\usepackage[version=4]{mhchem}
\pgfplotsset{compat=1.17}
\usepackage{subcaption}
\usetikzlibrary{shapes.geometric}
\usetikzlibrary{shadows,patterns,perspective}
\usetikzlibrary {decorations,decorations.text}
\usetikzlibrary{decorations.pathreplacing}
\usepackage{float}
\usepackage{hyperref}
\usepackage{xcolor}

\definecolor{linkcolor}{RGB}{0, 0, 255}      % Blue color for links
\definecolor{citecolor}{RGB}{0, 128, 0}     % Green color for citations
\definecolor{urlcolor}{RGB}{255, 0, 0}       % Red color for URLs

\hypersetup{
    colorlinks=true,
    linkcolor=linkcolor,
    citecolor=citecolor,
    urlcolor=urlcolor,
    linktoc=all,  % Set links in the table of contents to be colored as well
    pdfborder={0 0 0}  % Removes the box around links
}

\DeclareCaptionJustification{justified}{\leftskip=0pt \rightskip=0pt \parfillskip=0pt plus 1fil}

\begin{document}
\definecolor{dy}{rgb}{0.9,0.9,0.4}
\definecolor{dr}{rgb}{0.95,0.65,0.55}
\definecolor{db}{rgb}{0.5,0.8,0.9}
\definecolor{dg}{rgb}{0.2,0.9,0.6}
\definecolor{BrickRed}{rgb}{0.8,0.3,0.3}
\definecolor{Navy}{rgb}{0.2,0.2,0.6}
\definecolor{DarkGreen}{rgb}{0.1,0.4,0.1}

\title{The measurement-induced phase transition in strongly disordered spin chains}

\author{Yicheng Tang}
\affiliation{Department of Physics and Astronomy, Center for Materials Theory,
Rutgers University, Piscataway, NJ 08854, United States of America}
\author{Pradip Kattel}
\affiliation{Department of Physics and Astronomy, Center for Materials Theory,
Rutgers University, Piscataway, NJ 08854, United States of America}
\affiliation{Department of Quantum Matter Physics, University of Geneva, Quai Ernest-Ansermet 24, 1211 Geneva, Switzerland}
\author{Arijeet Pal}
\affiliation{Department of Physics and Astronomy, University College London,
Gower Street, London WC1E 6BT, United Kingdom}
\author{Emil A. Yuzbashyan}
\affiliation{Department of Physics and Astronomy, Center for Materials Theory,
Rutgers University, Piscataway, NJ 08854, United States of America}
\author{J. H. Pixley}
\affiliation{Department of Physics and Astronomy, Center for Materials Theory,
Rutgers University, Piscataway, NJ 08854, United States of America}
\affiliation{Center for Computational Quantum Physics, Flatiron Institute, 162 5th Avenue, New York, NY 10010}

\begin{abstract}
We investigate the dynamics of strongly disordered spin chains in the presence of random local measurements. By studying the transverse-field Ising model with a site-dependent random longitudinal field and an effective $l$-bit many-body localized Hamiltonian, we show that the prethermal and MBL regimes are unstable to   local measurements along any direction. Any non-zero measurement density induces a volume-law entangled phase with a subsequent  phase transition into an area-law  state as the measurement rate is further increased.
%due to measurements acting as non-commuting perturbations in the $l$-bit basis with finite support. 
The critical measurement rate $p_c$, where the transition occurs, is exponentially small in the strength of disorder  $W$ and the average overlap between the measurement operator and the local integrals of motion $O$ as $p_c \sim \exp[-\alpha W/(1-O^2)]$. In the measurement-induced volume-law phase, the saturation time scales as $t_s \sim L $, contrasting the exponentially slow saturation $t_s \sim e^{aL}$ in the prethermal and MBL regimes at $p = 0$.
% Time scales in the measurement-induced volume-law phase are analyzed and shown to scale linearly with the system size and not to connect smoothly with the MBL limit, where the time scale is exponentially long.
%$ %The critical properties are analyzed in this model and the $l$-bit limit finding reasonable agreement with the universallity class of the  MIPT in Haar random circuits.
\end{abstract}

\maketitle
% \section{Introduction}

Localization of quantum particles has remained of central interest in condensed matter physics~\cite{abrahams1979scaling,alet2018many}. While it is well known that strong disorder localizes noninteracting fermions~\cite{PhysRev.109.1492,anderson201050}, the role of interactions in many-body localization (MBL) at finite energy density is a subtle and challenging question~\cite{oganesyan2007localization,nandkishore2015many,abanin2019colloquium,geraedts2016many,imbrie2016many,vsuntajs2020quantum}. %The presence of a prethermal regime resembling MBL on small sizes precedes any putative transition to a fully MBL phase. 
The prethermal regime preceding the MBL phase exhibits dynamical properties resembling full MBL~\cite{RoeckPhysRevB.95.155129,long2023phenomenology,Abanin2017RMP}, for small system sizes, in classical and quantum simulations~\cite{luitz2015many,bardarson2012unbounded,PhysRevB.77.064426}. The instability of the MBL regime to quantum measurements can provide a new perspective on localization, as random local measurements induce a form of dynamical localization through wavefunction collapse, which is distinct from quantum interference.  In $d$-dimensional monitored free-fermionic circuits, this interplay has been made explicit by mapping the system to $d+1$-dimensional free fermions in quenched disorder; the Anderson localized phase in $d+1$ now corresponds to an area-law scaling of the entanglement entropy~\cite{li2018quantum,skinner2019measurement,Altman2021,Nahum2021PRXQ}. However, in the presence of strong interactions, this mapping no longer applies, and the interplay of disorder, interactions, and measurements offers a versatile setting to explore new dynamical effects and phase transitions.

\begin{figure}
    \centering    \includegraphics[width=\linewidth]{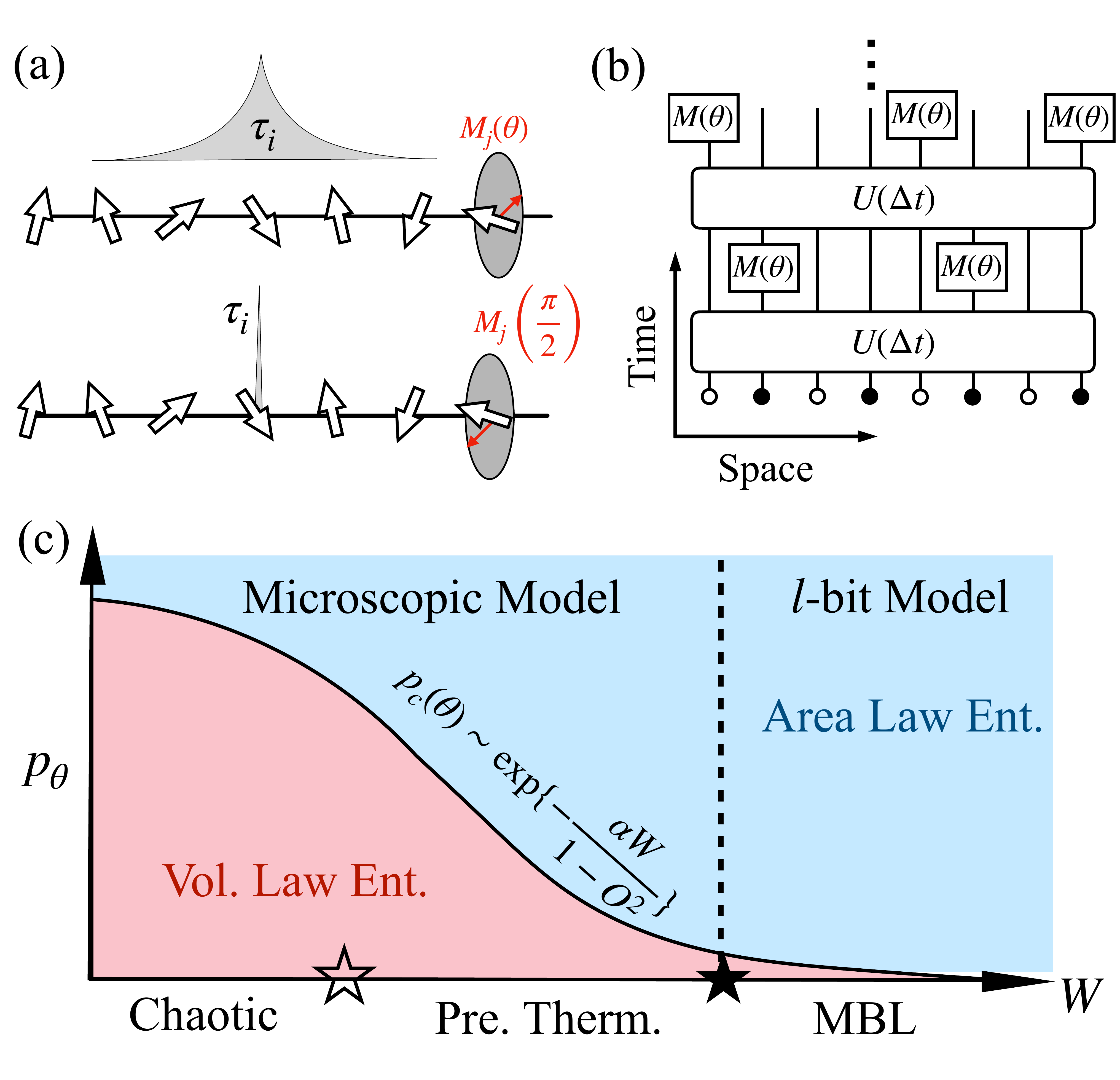}
    \caption{(a) The top panel is a schematic of the typical
    support of an $l$-bit operator, localized at site $i$ in the microscopic model, with each qubit measured randomly with $M_i(\theta)$ in the $x-z$ plane. The bottom panel shows a schematic of the $l$-bit operator in the $l$-bit model, with each qubit measured randomly in the $x$ direction.  
    (b) Illustration of measurement protocol with unitary Hamiltonian dynamics $U(\Delta t)$  followed by random measurements $M_i(\theta)$ with rate $p$. (c) Illustration of the  phase diagram for the MBL system with measurement in the space of disorder strength $W$ and measurement rate $p_{\theta}$ at angle $\theta$.
    For any $\theta$, the critical measurement rate $p_c\sim \exp\{- \frac{\alpha W}{1-O^2}\}$ for large $W$ labels the boundary between volume and area-law entangled phases with $O$ being the overlap between the measurement operator (defined eplicitly for in Eq.~\eqref{Overlap}).}
    \label{Fig1}
\end{figure}

Surprisingly, in the presence of interactions, localization induced by static disorder and measurement are, in fact, directly competing as they correspond to different forms of localized behavior. %Previous study of the continuous measurement \cite{PhysRevB.107.L220201} and the no-click limit \cite {PhysRevB.109.174205} indicates MBL under the measurement along the disorder direction is stable. 
This is due to the physical spin operators having finite support over many $l$-bits, each projective measurement acts as a non-commuting perturbation on an extended set of integrals of motion. Repeated measurements, therefore, continually scramble the $l$-bit degrees of freedom and destabilize the MBL structure, even when the overlap with a single $l$-bit is exponentially small, which was studied by analogous stabilizer circuits mapped to measurement-only models~\cite{Ippoliti2021PRX}.
% Applying local measurements to a model in the MBL phase (that is represented via a series $l$-bits serving as quasi-local conservation laws) {\it generates} entanglement, through the MBL unitaries ``stretching'' each measurement operator across several sites. Non-local and non-commuting measurements can realize a volume-law phase for finite measurement rates analogous to measurement-only models~\cite{Ippoliti2021PRX}.
At sufficiently high measurement rates (in particular directions), the MBL model has been found to undergo a measurement-induced phase transition (MIPT) into an area-law phase, which is unrelated to the underlying MBL phase. In contrast, the volume-law phase ceases to exist for measurements of the exact $l$-bit operator~\cite{PhysRevResearch.2.043072}, which are still exponentially complex to construct.  In practice, only physical spins, which have partial overlaps with $l$-bits, are measurable, as depicted in Fig.~\ref{Fig1}(a). This deviation plays a crucial non-perturbative role in the instability of MBL dynamics to measurements. 
A precise understanding of how measurements impact the stability of the prethermal and MBL regimes has been lacking and is crucial for a comprehensive picture of the instabilities of MBL in open system dynamics. In particular, prior work~\cite{PhysRevResearch.2.043072,PhysRevB.107.L220201,PhysRevB.109.174205} concluded that if you measure physical operators that ``align'' strongly with the $l$-bit operators then there is no volume law phase. However, we show that this conclusion is incorrect and provide strong numerical evidence of a non-zero critical measurement rate  that follows a universal form for any measurement direction that is albeit exponentially small in the disorder strength, thus demonstrating the instability of the MBL phase to random local measurements.

In this work, we investigate the monitored quantum dynamics of  strongly disordered spin-1/2 chains without any space-time symmetries other than the energy conservation. We consider their unitary Hamiltonian dynamics  interspersed by random local measurements along an arbitrary direction denoted by the angle $\theta$, see Fig.~\ref{Fig1}. To compute the dynamics, we utilize a Chebyshev expansion of the unitary time evolution operator~\cite{RevModPhys.78.275}, which allows an efficient and accurate truncation. As a result, we achieve system sizes comparable to those of quantum circuit simulations, which help us to determine the phase boundary separating volume-law and area-law entangled phases. 

We show that the boundary is exponentially small in both the disorder strength and the overlap of local measurement operators with the $l$-bit degrees of freedom. 
We complement this analysis with a study of the saturation time scales in the $l$-bit limit of the model, which shows that the $p>0$ region of the phase diagram, where this scale grows linearly with the system size,  is not smoothly connected to the $p=0$ line, where it is exponentially slow. These results are supported by both the spin-chain model in the prethermal MBL regime and the $l$-bit Hamiltonian, suggesting that they are generic, and can be tested on current quantum platforms such as trapped-ion and superconducting-qubit experiments that have realized MBL~\cite{Smith2016,Choi2016,kucsko2018critical,gong2021experimental,roushan2017spectroscopic,chiaro2020growth,kohlert2019observation,zhu2021probing,schreiber2015observation,bordia2017probing} and MIPTs~\cite{noel2022measurement,koh2022experimental,google2023measurement,agrawal2024observing,feng2025postselection}. 

\begin{figure*}
    \centering
    \includegraphics[width=\linewidth]{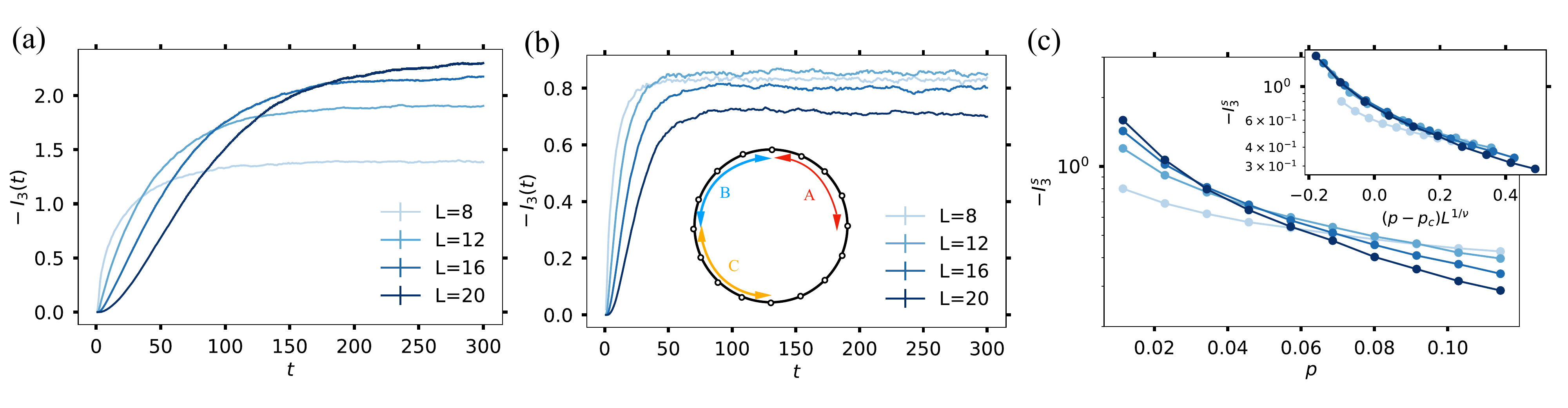}
    \caption{Tripartite mutual information $I_3(t)$ for the evolution under Hamiltonian~\eqref{Hmbl} and random measurements for measurement  angle $\theta = \frac{\pi}{2}$ and  various system sizes $L$. The   disorder strength is $W=5$ and  the measurement probability is $p=0.01$ in (a) and  and $p=0.09$ in (b) corresponding to the volume-law and the area-law entangled phases, respectively. The inset in (b) depicts  the geometry used to compute $I_3(t)$.  (c) Tripartite mutual information in the steady state $I_3^s$ as a function of $p$ with a critical value $p_c\simeq 0.039$. The inset shows  the data collapse for the microscopic model using   the critical exponent $\nu\simeq 1.6$ extrapolated from the $l$-bit model.}
    % {\color{red} JP: 1) I am confused by this last sentence. What are you saying? I thought you can collapse this reasonable well in some cases? 2) If this data is all for Eq. 1 model state that in the first sentence of the caption}
    % }
    \label{fig:I3t}
\end{figure*}

% \section{Models and Approach}
To study the interplay of measurements and MBL, we focus on a repeated two-stage protocol, which consists of alternating unitary evolution and projective measurements as shown in Fig.~\ref{Fig1}(b).
In the unitary step, the system evolves under the unitary operator that we take to be the same each time it is applied (i.e., it is a Floquet operator) $U=e^{-iH\Delta t}$, and we set $\Delta t  = 1/J$ throughout. We consider two different Hamiltonians in this work: the microscopic Hamiltonian~\eqref{Hmbl} and the $l$-bit Hamiltonian~\eqref{Hlbit}.

% \subsection{Microscopic Hamiltonian}
The microscopic Hamiltonian  that we analyze is the disordered transverse-field Ising model with a site-dependent random longitudinal field
\begin{equation}
    H(W,\{\eta\}) = J\sum_{i=1}^L\left( \sigma^z_i\sigma^z_{i+1}+\sigma^x_{i}+W\eta_i\sigma^z_i\right).
    \label{Hmbl}
\end{equation}
The strength of disorder is denoted $W,$ and $\eta_i$ are random variables at each site that are drawn from the uniform distribution between $[-1,1]$.  
To ascertain the small-$L$ behavior of this Hamiltonian, we analyze its level spacing ratio using exact diagonalization and present the results in End matter. We observe a clear transition from Wigner-Dyson to Poisson-like behavior as the disorder is increased, with a crossing that drifts up to $W\gtrsim 3.2$, which has been interpreted as the onset of MBL in small systems. However, recent studies suggest that true many-body localization may only occur for significantly stronger disorder strengths in the thermodynamic limit ~\cite{PhysRevB.106.L020202,PhysRevB.105.174205,long2023phenomenology}. In the intermediate prethermal regime $W\gtrsim 3.2$, the system displays slow entanglement dynamics and suppressed  thermalization, which is the parameter regime of the microscopic model we focus on in this work.

%In an MBL system, the Hamiltonian commutes with a set of emergent local integrals of motion ($l$-bits) $\tau^z_i$, which satisfy the Pauli algebra and can be expressed in terms of the physical qubit operators $\sigma$s as~\cite{serbyn2013local,huse2014phenomenology} {\bf (CITATIONS to original $l$-bit papers)}
%\begin{equation}
%    \tau^z_i = \sum_{i,\alpha}t_i^\alpha\sigma^\alpha_i +\sum_{i,j,\alpha,\beta}t_{ij}^{\alpha\beta}e^{-m|i-j|}\sigma^\alpha_i\sigma^\beta_j+...
%\end{equation}

%The critical MBL randomness strength $W$ for the Hamiltonian defined in Eq.~\eqref{Hmbl} is known to exhibit strong finite-size corrections, implying that for the $W$ values considered in the previous section, the system may remain in a prethermal regime rather than having fully entered the MBL phase. 
% \subsection{$l$-bit Hamiltonian}
In addition to analyzing  the prethermal regime of this finite system size MBL model, we also investigate another Hamiltonian (the $l$-bit Hamiltonian), which is by construction in the MBL phase.
When the system is in the MBL phase,  there exists a complete set of emergent local integrals of motion, illustrated in Fig.~\ref{Fig1}(a).
Therefore, to study the effects of measurement in the MBL phase, we consider an effective MBL Hamiltonian in the $l$-bit ($\tau$) basis~\cite{serbyn2013local,huse2014phenomenology}, which is related to the physical $\sigma$ basis by a local unitary transformation. In this basis, the diagonal form of the MBL Hamiltonian is
\begin{equation}
\label{Hlbit}
    H_{l-\mathrm{bits}}(\{\eta\}) = \sum_{i} h\eta_i\tau^z_i+\sum_{i,j}\eta_{ij}e^{-m|i-j|}\tau^z_i\tau^z_j+...
\end{equation}
where $...$ stands for the higher order terms in terms of the $l$-bits, $\eta\in[0,1]$ is a random variable. In the following, unless otherwise specified, we drop the higher-order terms and choose $h=5$ for the $l$-bit model.  Without measurement, $p=0$, the Hamiltonian dynamics generates entanglement entropy with a logarithmic growth $S(t)\sim \ln t$ characteristic of MBL systems~\cite{PhysRevLett.110.260601,PhysRevB.77.064426,bardarson2012unbounded} for both the microscopic Hamiltonian~\eqref{Hmbl} and the $l$-bit Hamiltonian~\eqref{Hlbit}.

In the measurement step, we perform a projective measurement with probability $p$ at each site.  We measure in the $xz$ plane at site $i$  the operator 
\begin{equation}
    M_i(\theta) = \cos\theta \sigma^z_i+\sin\theta \sigma^x_i,
    \label{eq:theta}
\end{equation}  which interpolates between $\sigma^z_i = M_i (0)$ and $\sigma^x_i=M_i\left(\frac{\pi}{2}\right)$ and has measurement outcome $m_i=\pm 1$. This measurement  consists of the projectors  $P_{m_i}(\theta)=(1\pm M_i(\theta))/2$ with eigenvalues $m_i=\pm 1$ and the state transforms according to
\begin{equation}
    |\psi\rangle \rightarrow \frac{1}{p_{m_i}}P_{m_i}(\theta)|\psi\rangle,
\end{equation}
with Born probability $p_{m_i} = \langle \psi| P_{m_i}(\theta)|\psi\rangle$ up to normalization.
As a result, the time evolution is non-unitary, and the state is now indexed by the measurement history labeled by the set $\{m_k\}$. 
Notice  that purely $\tau^z$ measurements in the $l$-bit model commute with the Hamiltonian and produce a steady state that is a product state for any finite measurement rate. Therefore, only measurements with a finite $\tau_x$ component lead to nontrivial dynamics.

Combing the unitary dynamics with the above measurement protocol,  we evolve the system  by iterating these two steps as shown in Fig.~\ref{Fig1}(b), starting from a random product state  $\ket{\psi_0}=\otimes _{i=1}^L\ket{\theta_i,\phi_i}$ where each qubit is initialized in a pure state parameterized by spherical angles $\theta_i$ and $\phi_i$ as $\ket{\theta,\phi}=(\cos\frac{\theta}{2},e^{i\phi}\sin\frac{\theta}{2})$. At every time step $t$, we first evolve the state $|\psi(t)\rangle$ as  $|\psi(t+\frac{1}{2})\rangle = e^{-iH(\{\eta\}\Delta t}|\psi(t)\rangle$. To simulate the time evolution of the system under Hamiltonian dynamics, we approximate the unitary time-evolution operator with a Chebyshev polynomial expansion, as discussed  in End matter. A benefit of the Hamiltonian acting over a short time for each step of the evolution is that it allows us to truncate the Chebyshev expansion at a low order. After the unitary evolution, the measurement step is applied, $|\psi(t+1)\rangle = \prod_{\{m_i\}}\frac{1}{p_{m_i}}P_{m_i}|\psi(t+\frac{1}{2})\rangle$ with a set of measurement outcomes $m_i(t)$ if the $i$-th qubit is measured. The resulting state at time $t$ depends on the random variable $\{\eta\}$ and measurement outcomes $\{m_i(t)\}$, and we label it as $\ket{\psi(t)}_{\{\eta,m_i(t)\}}$.

%on the intial condition $\{\theta,\phi\}$

At each step $t$, we compute the tripartite mutual information for the state $\ket{\psi(t)}_{\{\eta,m_i(t)\}}$ defined as
$I^{\{\eta,m_i(t)\}}_3(t) = S_{A}+S_{B}+S_C+S_{ABC}-S_{AB}-S_{BC}-S_{AC}$,
where $S_X = -\rho_X\ln \rho_X$ is the von-Neumann entropy of subsystem $X$ with reduced density matrix $\rho_X=\mathrm{Tr}_{\bar{X}}\left(\ket{\psi(t)}_{\{\eta,m_i(t)\}}\bra{\psi(t)}_{\{\eta,m_i(t)\}}\right)$. The subsystems $A$, $B$, and $C$ are shown in Fig.~\ref{fig:I3t}(b); each corresponds to one quarter of the system.
We average the mutual information over $2000$ samples of random parameters $\{\eta\}$ and quantum trajectories $\{m_i(t)\}$, $I_3(t)=\bigl[\bar{I}^{\{\eta,m_i(t)\}}_3(t)\bigr]$, with $\bar{X}$ indicating the average of $X$ over random parameters $\{\eta\}$  and $[X]$ indicating the average over the quantum trajectory $\{m_i(t)\}$. 
% {\color{red} JP: See above, define properly as seperate disorder averages over $P[\eta]$ and one over measurement outcomes weighted by the Born probability.}% \section{Results}
% \subsection{Microscopic Hamiltonian}
In Fig.~\ref{fig:I3t} (a) and (b), we present two representative examples of the time evolution of $I_3(t)$ with parameters 
$W=5$ and $\theta=\tfrac{\pi}{2}$, differing only in the value of the measurement rate $p$. Both show
 saturation to a steady-state value $I_3^s=I_3(t\to\infty)$ under the repeated measurement protocol.

To probe the entanglement structure of the steady state, we analyze the scaling of the steady state tripartite mutual information  $I_3^s$ as a function of the system size $L$ and measurement rate $p$ at a fixed measurement angle $\theta=\pi/2$ and randomness $W=5$, see Fig.~\ref{fig:I3t} (c). 
A transition between volume-law and area-law entanglement is detected from the crossing of the $I_3^s(p)$ curves for different system sizes $L$, which determines the critical measurement rate $p_c$. Near the critical point in the long time limit, the tripartite mutual information follows a universal scaling function~\cite{PhysRevB.101.060301,Bao2020PRB, Gullans2020PRL}
\begin{equation}
    I_3^s=I_3(t\rightarrow \infty)\sim f\left((p-p_c)L^{1/\nu}\right)
    \label{Eq.collapse}
\end{equation}
where $\nu$ is the correlation length exponent and $f(x)$ is a smooth function. Because of  strong finite-size effects, $\nu$ cannot be reliably extracted for the microscopic model. However, the $l$-bit model yields $\nu=1.6\pm0.01$.
%all data collapse onto a universal curve when plotted against the rescaled variable $(p-p_c)L^{1/\nu}$ \cite{PhysRevB.101.060301}, yielding a critical exponent $\nu \simeq 1.3$, consistent with the value found in the MIPT of Haar random circuits.

\begin{figure}
    \centering
    \includegraphics[width=\linewidth]{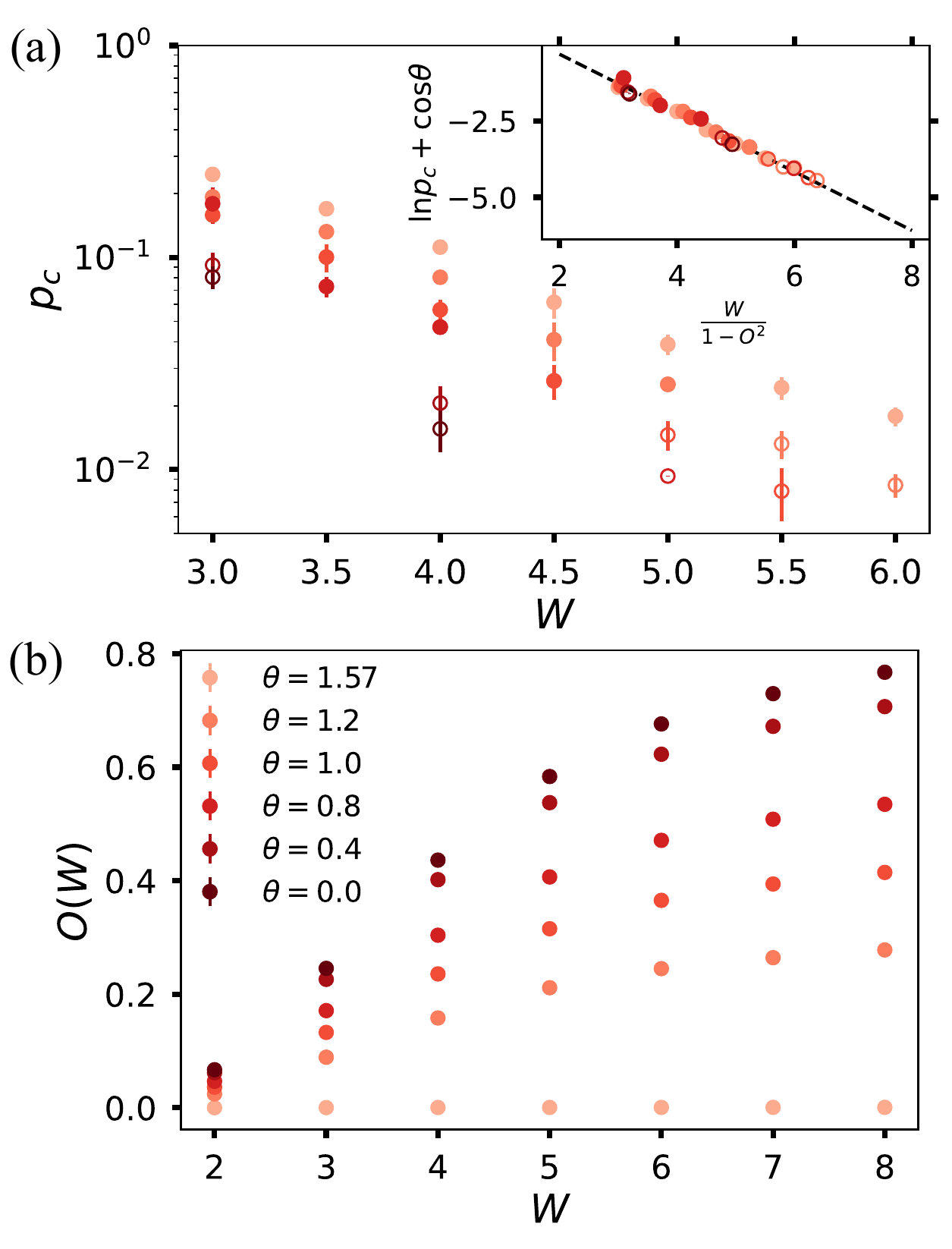}
    \caption{(a) The $W$- and $\theta$-dependence of the critical measurement rate $p_c(W,\theta)$ defines the phase boundary between volume-law entanglement, $p<p_c$, and area-law entanglement, $p>p_c$, for the model~\eqref{Hmbl}.  The inset shows the data collapse onto the curve given by Eq.~\eqref{eq_pc}. The hollow markers indicate data points where $p_c$ shows a significant finite-size effect. (b) Overlaps between the local integral of motion and the measurement operator $O(\theta,W)$ as defined in 
    Eq.~\eqref{Overlap} for Hamiltonian~\eqref{Hmbl}}
    \label{fig:p_c}
\end{figure}

\begin{figure*}
    \centering \includegraphics[width=\linewidth]{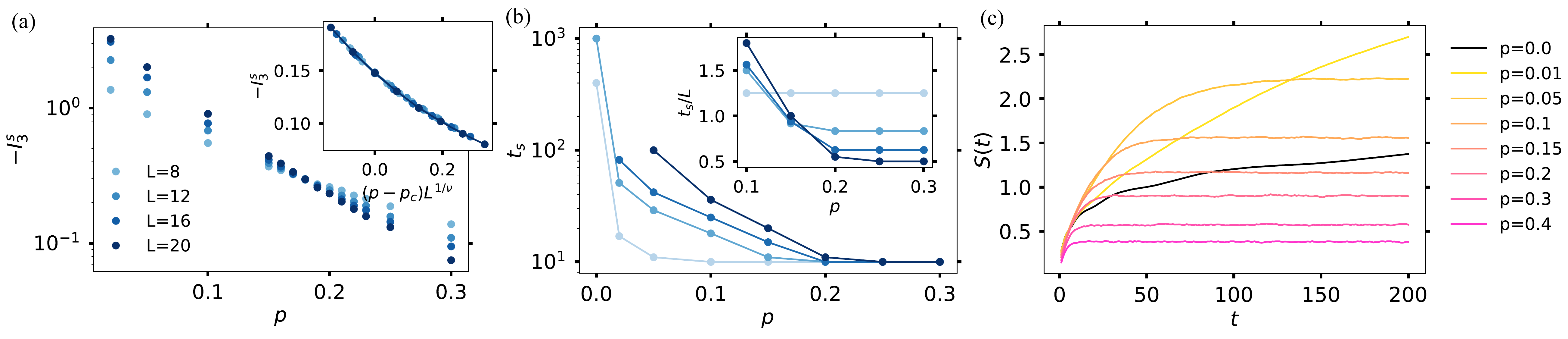}
    \caption{Entanglement dynamics for the $l$-bit model \eqref{Hlbit}. (a) Plots of the steady state mutual information $I_3^s$ vs. the measurement rate $p$ for different system sizes $L$ crossing  at $p=p_c$. The inset shows the  data collapse with the scaling $(p-p_c)L^{1/\nu}$ as defined in Eq.~\eqref{Eq.collapse}, where $\nu\simeq 1.6$. (b) Saturation time $t_s$ ($t_s/L$ in the inset) as a function of $p$. The $t_s/L$ plots collapse consistent with $t_s\sim\mathcal{O}(L)$ when $p<p_c=0.18$ and the system is in the volume-law entanglement phase, while the $t_s$ plots collapse for $p>p_c$ when the system is in the area-law entangled phase where $t_s\sim\mathcal{O}(1)$. Legend is shared with (a). (c) Half-cut entanglement entropy for different measurement rates $p$ demonstrating that  the entanglement entropy can grow faster with measurements than without.}
    \label{fig:Lbit}
\end{figure*}

The complete phase diagram  is obtained by extracting the critical measurement rate $p_c(\theta, W)$
from the crossings of the $I_3^s(p)$ curves at various values of $W$ and $\theta$. We observe that for large randomness $W\geq 3$, the critical measurement rate is exponentially small $p_c\sim e^{-f(\theta)W}$ as shown in Fig.~\ref{fig:p_c}. To determine  $f(\theta)$, we compute the overlap between the measurement operator in Eq.~\eqref{eq:theta} and the emergent local integral of motion (per disorder sample), defined as ~\cite{chandran2015constructing}
\begin{equation}
    O(\theta,W) = \frac{1}{2^L}\mathrm{Tr}\left(M(\theta)\bar{L}_i(W,\{\eta\})\right),
    \label{Overlap}
\end{equation}
where $L_i(W,\{\eta\}) = V^\dagger \mathrm{diag}(V\sigma^z_iV^\dagger)V$ and $V$ is the unitary operator that diagonalizes the Hamiltonian $H(W,\{\eta\})$ and $\mathrm{diag}(A)$ is a diagonal matrix with the same diagonal elements as $A$.
By construction, the local integral of motion $L_i(W,\{\eta\})$ is a conserved charge, $[L_i,H]=0$, which  depends only on the Hamiltonian. The numerical values of $O(\theta,W)$ averaged over 500 samples of random variables $\eta_i$   are shown in 
Fig.~\ref{fig:p_c}(b). 

This overlap characterizes the extent to which measurement ``freezes'' an integral of motion. We compare the critical measurement rate $p_c$ with the overlap $O$ shown in Fig.~\ref{fig:p_c} with an ansatz 
\begin{equation}
    p_c(W,\theta) = 
    %p_0 e^{-\cos\theta}\exp\left[-\frac{\alpha W}{1-O(\theta,W)^2}\right]
    p_0 \exp\left(-\frac{\alpha W}{1-O(\theta,W)^2}-\cos\theta\right)
    \label{eq_pc}
\end{equation}
where $p_0=4.9$ and $\alpha=0.96$ are obtained by collapsing the data onto a single curve, see the inset in Fig.~\ref{fig:p_c}. 
Eq.~\eqref{eq_pc} implies several key consequences. First,  it indicates that there is always a phase transition from volume to area-law entanglement in the prethermal and MBL systems due to measurements for any value of the overlap between the measurement operator and the $l$-bit degrees of freedom. 
At the same, the phase boundary out of the volume-law phase is exponentially small as the parameters $W$ and $\theta$ are increased.
Finally, it also suggests that the prethermal MBL regime is unstable to any non-zero measurement rate  in any direction, irrespective of the nature of the $l$-bit operators. To understand this instability, we now turn to the $l$-bit model~\eqref{Hlbit}.

% \subsection{$l$-bit Hamiltonian}
The $l$-bit Hamiltonian puts us deep into the MBL phase by construction, and we find that the   finite-size effects are significantly suppressed  as compared to the microscopic model~\eqref{Hlbit} because the localization length is strictly bounded. 
This allows us to obtain better numerical results for $I_3^s$ and establish that the $\tau^x$ measurements produce an MIPT with $p_c=0.18$ and the critical exponent $\nu\approx 1.6\pm 0.01$ as shown in Fig.~\ref{fig:Lbit}(a). Moreover, the dynamical exponent $z=1$ as we  demonstrate in the End Matter.

In Fig.~\ref{fig:Lbit}(c), we show that in the presence of measurements, the entanglement entropy grows faster than in the case of purely unitary dynamics destroying the logarithmic growth in the MBL state and indicating that the MBL dynamics is unstable with respect to any non-zero density of local measurements.  To study the instability of the MBL phase under measurements, we investigate the system at the saturation time $t_s$, defined by $I_3(t>t_s)=I_3^s,$ see Fig.~\ref{fig:Lbit}(b). In the MBL phase ($p=0)$, $t_s$ scales exponentially with the system size $ t_s \sim e^{aL}$~\cite{bardarson2012unbounded}. However, when a finite density of measurements is present ($p>0$), 
the exponentially slow relaxation collapses to a linear scaling $t_s \sim L$ characteristic of a volume-law entangled phase. Throughout the volume-law regime $(0<p < p_c)$ one finds $t_s \propto L$. By contrast, in the area-law phase $(p > p_c)$ the saturation time becomes size-independent, $t_s \sim \mathcal{O}(1)$.

% \section{Conlusion}
% We discussed the dynamics of MBL systems under the effect of local single-qubit measurements. We showed that the MBL Hamiltonian defined in Eq.\eqref{Hmbl} has a MIPT under arbitrary single-qubit measurement, with a critical measurement rate that decreases exponentially as the randomness $W$ increases. To mitigate the strong finite-size effect in the MBL system, we examined the measurement effect in the $l$-bit model, which, by construction, is in the MBL phase. The numerical results suggest that the $l$-bit model also undergoes an MIPT as the measurement rate increases.

We resolve the phase diagram of monitored MBL systems by determining the disorder-dependent critical measurement rate $p_{c}$ at which the entanglement structure changes qualitatively. Starting from an MBL regime with slow, logarithmic entanglement growth, we find that any nonzero measurement rate immediately produces a stable volume-law phase. Upon further increasing the rate, the system undergoes a sharp transition at a finite $p_{c}(W,\theta)$ into a measurement-dominated area-law phase, with $p_{c}$ decreasing exponentially with disorder strength. By mapping $p_{c}(W,\theta)$ across disorder and measurement angles, we completely determine the  boundary between these phases. The resulting transition is a property of the monitored dynamics itself and does not track the underlying unitary MBL crossover, demonstrating that MBL systems, once monitored, generically exhibit a three-stage structure: slow-entangling MBL at $p=0$, a robust volume-law regime for any $p_c(W,\theta)>p>0$, and an area-law phase for $p>p_{c}(W,\theta)$.

{\it Acknowledgement}:
We thank Anushya Chandran, Vedika Khemani, and Sarang Gopalakrishnan for useful discussions, and we thank David Huse for in-depth discussions. This work is partially supported by the Army Research Office Grant No.~W911NF-23-1-0144 (P.K.~and J.H.P.) and the Rutgers Samuel
Marateck Fellowship (Y.T.). A.P. was funded by the European Research Council (ERC) under the EU’s Horizon 2020 research and innovation program (Grant Agreement No.~853368). A.P. thanks the Physics departments at Princeton and Harvard Universities for their support and hospitality during the sabbatical where part of this work was carried out. 

\bibliography{ref}

@article{PhysRevB.107.L220201,
  title = {Localization properties in disordered quantum many-body dynamics under continuous measurement},
  author = {Yamamoto, Kazuki and Hamazaki, Ryusuke},
  journal = {Phys. Rev. B},
  volume = {107},
  issue = {22},
  pages = {L220201},
  numpages = {7},
  year = {2023},
  month = {Jun},
  publisher = {American Physical Society},
  doi = {10.1103/PhysRevB.107.L220201},
  url = {https://link.aps.org/doi/10.1103/PhysRevB.107.L220201}
}

@article{PhysRevB.109.174205,
  title = {Stable many-body localization under random continuous measurements in the no-click limit},
  author = {De Tomasi, Giuseppe and Khaymovich, Ivan M.},
  journal = {Phys. Rev. B},
  volume = {109},
  issue = {17},
  pages = {174205},
  numpages = {12},
  year = {2024},
  month = {May},
  publisher = {American Physical Society},
  doi = {10.1103/PhysRevB.109.174205},
  url = {https://link.aps.org/doi/10.1103/PhysRevB.109.174205}
}

@article{PhysRevB.77.064426,
  title = {Many-body localization in the Heisenberg $XXZ$ magnet in a random field},
  author = {\ifmmode \check{Z}\else \v{Z}\fi{}nidari\ifmmode \check{c}\else \v{c}\fi{}, Marko and Prosen, Toma\ifmmode \check{z}\else \v{z}\fi{} and Prelov\ifmmode \check{s}\else \v{s}\fi{}ek, Peter},
  journal = {Phys. Rev. B},
  volume = {77},
  issue = {6},
  pages = {064426},
  numpages = {5},
  year = {2008},
  month = {Feb},
  publisher = {American Physical Society},
  doi = {10.1103/PhysRevB.77.064426},
  url = {https://link.aps.org/doi/10.1103/PhysRevB.77.064426}
}

@article{PhysRevLett.110.260601,
  title = {Universal Slow Growth of Entanglement in Interacting Strongly Disordered Systems},
  author = {Serbyn, Maksym and Papi\ifmmode \acute{c}\else \'{c}\fi{}, Z. and Abanin, Dmitry A.},
  journal = {Phys. Rev. Lett.},
  volume = {110},
  issue = {26},
  pages = {260601},
  numpages = {5},
  year = {2013},
  month = {Jun},
  publisher = {American Physical Society},
  doi = {10.1103/PhysRevLett.110.260601},
  url = {https://link.aps.org/doi/10.1103/PhysRevLett.110.260601}
}

@article{PhysRevResearch.2.043072,
  title = {Measurement-induced entanglement transitions in many-body localized systems},
  author = {Lunt, Oliver and Pal, Arijeet},
  journal = {Phys. Rev. Res.},
  volume = {2},
  issue = {4},
  pages = {043072},
  numpages = {10},
  year = {2020},
  month = {Oct},
  publisher = {American Physical Society},
  doi = {10.1103/PhysRevResearch.2.043072},
  url = {https://link.aps.org/doi/10.1103/PhysRevResearch.2.043072}
}

@book{fehske2007computational,
  title={Computational many-particle physics},
  author={Fehske, Holger and Schneider, Ralf and Weisse, Alexander},
  volume={739},
  year={2007},
  publisher={Springer}
}

@article{PhysRevB.101.060301,
  title = {Critical properties of the measurement-induced transition in random quantum circuits},
  author = {Zabalo, Aidan and Gullans, Michael J. and Wilson, Justin H. and Gopalakrishnan, Sarang and Huse, David A. and Pixley, J. H.},
  journal = {Phys. Rev. B},
  volume = {101},
  issue = {6},
  pages = {060301},
  numpages = {5},
  year = {2020},
  month = {Feb},
  publisher = {American Physical Society},
  doi = {10.1103/PhysRevB.101.060301},
  url = {https://link.aps.org/doi/10.1103/PhysRevB.101.060301}
}

@article{vsuntajs2020quantum,
  title={Quantum chaos challenges many-body localization},
  author={{\v{S}}untajs, Jan and Bon{\v{c}}a, Janez and Prosen, Toma{\v{z}} and Vidmar, Lev},
  journal={Physical Review E},
  volume={102},
  number={6},
  pages={062144},
  year={2020},
  publisher={APS}
}

@article{bardarson2012unbounded,
  title={Unbounded growth of entanglement in models of many-body localization},
  author={Bardarson, Jens H and Pollmann, Frank and Moore, Joel E},
  journal={Physical review letters},
  volume={109},
  number={1},
  pages={017202},
  year={2012},
  publisher={APS}
}

@article{skinner2019measurement,
  title={Measurement-induced phase transitions in the dynamics of entanglement},
  author={Skinner, Brian and Ruhman, Jonathan and Nahum, Adam},
  journal={Physical Review X},
  volume={9},
  number={3},
  pages={031009},
  year={2019},
  publisher={APS}
}

@article{long2023phenomenology,
  title={Phenomenology of the prethermal many-body localized regime},
  author={Long, David M and Crowley, Philip JD and Khemani, Vedika and Chandran, Anushya},
  journal={Physical Review Letters},
  volume={131},
  number={10},
  pages={106301},
  year={2023},
  publisher={APS}
}

@article{PhysRevB.106.L020202,
  title = {Bath-induced delocalization in interacting disordered spin chains},
  author = {Sels, Dries},
  journal = {Phys. Rev. B},
  volume = {106},
  issue = {2},
  pages = {L020202},
  numpages = {5},
  year = {2022},
  month = {Jul},
  publisher = {American Physical Society},
  doi = {10.1103/PhysRevB.106.L020202},
  url = {https://link.aps.org/doi/10.1103/PhysRevB.106.L020202}
}

@article{PhysRevB.105.174205,
  title = {Avalanches and many-body resonances in many-body localized systems},
  author = {Morningstar, Alan and Colmenarez, Luis and Khemani, Vedika and Luitz, David J. and Huse, David A.},
  journal = {Phys. Rev. B},
  volume = {105},
  issue = {17},
  pages = {174205},
  numpages = {20},
  year = {2022},
  month = {May},
  publisher = {American Physical Society},
  doi = {10.1103/PhysRevB.105.174205},
  url = {https://link.aps.org/doi/10.1103/PhysRevB.105.174205}
}

@article{alet2018many,
  title={Many-body localization: An introduction and selected topics},
  author={Alet, Fabien and Laflorencie, Nicolas},
  journal={Comptes Rendus Physique},
  volume={19},
  number={6},
  pages={498--525},
  year={2018},
  publisher={Elsevier},
doi = {https://doi.org/10.1016/j.crhy.2018.03.003},
url = {https://www.sciencedirect.com/science/article/pii/S163107051830032X}
}

@article{chandran2015constructing,
  title={Constructing local integrals of motion in the many-body localized phase},
  author={Chandran, Anushya and Kim, Isaac H and Vidal, Guifre and Abanin, Dmitry A},
  journal={Physical Review B},
  volume={91},
  number={8},
  pages={085425},
  year={2015},
  publisher={APS}
}

@article{PhysRev.109.1492,
  title = {Absence of Diffusion in Certain Random Lattices},
  author = {Anderson, P. W.},
  journal = {Phys. Rev.},
  volume = {109},
  issue = {5},
  pages = {1492--1505},
  numpages = {0},
  year = {1958},
  month = {Mar},
  publisher = {American Physical Society},
  doi = {10.1103/PhysRev.109.1492},
  url = {https://link.aps.org/doi/10.1103/PhysRev.109.1492}
}

@article{anderson201050,
  title={50 Years of Anderson Localization},
  author={Anderson, PW},
  journal={50 Years Of Anderson Localization. Edited by PW Anderson. Published by World Scientific Publishing Co. Pte. Ltd},
  year={2010}
}

@article{abanin2019colloquium,
  title={Colloquium: Many-body localization, thermalization, and entanglement},
  author={Abanin, Dmitry A and Altman, Ehud and Bloch, Immanuel and Serbyn, Maksym},
  journal={Reviews of Modern Physics},
  volume={91},
  number={2},
  pages={021001},
  year={2019},
  publisher={APS}
}

@article{geraedts2016many,
  title={Many-body localization and thermalization: Insights from the entanglement spectrum},
  author={Geraedts, Scott D and Nandkishore, Rahul and Regnault, Nicolas},
  journal={Physical Review B},
  volume={93},
  number={17},
  pages={174202},
  year={2016},
  publisher={APS},
url={https://journals.aps.org/prb/abstract/10.1103/PhysRevB.93.174202}
}

@article{serbyn2013local,
  title={Local conservation laws and the structure of the many-body localized states},
  author={Serbyn, Maksym and Papi{\'c}, Zlatko and Abanin, Dmitry A},
  journal={Physical review letters},
  volume={111},
  number={12},
  pages={127201},
  year={2013},
  publisher={APS}
}

@article{huse2014phenomenology,
  title={Phenomenology of fully many-body-localized systems},
  author={Huse, David A and Nandkishore, Rahul and Oganesyan, Vadim},
  journal={Physical Review B},
  volume={90},
  number={17},
  pages={174202},
  year={2014},
  publisher={APS}
}

@article{abrahams1979scaling,
  title={Scaling theory of localization: Absence of quantum diffusion in two dimensions},
  author={Abrahams, Elihu and Anderson, Philip W and Licciardello, Donald C and Ramakrishnan, Tiruppattur V},
  journal={Physical Review Letters},
  volume={42},
  number={10},
  pages={673},
  year={1979},
  publisher={APS},
url={https://journals.aps.org/prl/abstract/10.1103/PhysRevLett.42.673}
}

@article{oganesyan2007localization,
  title={Localization of interacting fermions at high temperature},
  author={Oganesyan, Vadim and Huse, David A},
  journal={Physical Review B—Condensed Matter and Materials Physics},
  volume={75},
  number={15},
  pages={155111},
  year={2007},
  publisher={APS},
url={https://journals.aps.org/prb/abstract/10.1103/PhysRevB.75.155111}
}

@article{RoeckPhysRevB.95.155129,
  title = {Stability and instability towards delocalization in many-body localization systems},
  author = {De Roeck, Wojciech and Huveneers, Fran\ifmmode \mbox{\c{c}}\else \c{c}\fi{}ois},
  journal = {Phys. Rev. B},
  volume = {95},
  issue = {15},
  pages = {155129},
  numpages = {14},
  year = {2017},
  month = {Apr},
  publisher = {American Physical Society},
  doi = {10.1103/PhysRevB.95.155129},
  url = {https://link.aps.org/doi/10.1103/PhysRevB.95.155129}
}

@article{luitz2015many,
  title={Many-body localization edge in the random-field Heisenberg chain},
  author={Luitz, David J and Laflorencie, Nicolas and Alet, Fabien},
  journal={Physical Review B},
  volume={91},
  number={8},
  pages={081103},
  year={2015},
  publisher={APS}
}

@article{Smith2016,
  title     = {Many-body localization in a quantum simulator with programmable random disorder},
  author    = {Smith, J. and Lee, A. and Richerme, P. and Neyenhuis, B. and Hess, P. W. and Hauke, P. and Heyl, M. and Huse, D. A. and Monroe, C.},
  journal   = {Nature Physics},
  volume    = {12},
  pages     = {907--911},
  year      = {2016},
  doi       = {10.1038/nphys3783}
}

@article{Choi2016,
  title     = {Exploring the many-body localization transition in two dimensions},
  author    = {Choi, Jae-yoon and Hild, Sebastian and Zeiher, Johannes and Schau{\ss}, Peter and Rubio-Abadal, Antonio and Yefsah, Tarik and Khemani, Vedika and Huse, David A. and Bloch, Immanuel and Gross, Christian},
  journal   = {Science},
  volume    = {352},
  number    = {6293},
  pages     = {1547--1552},
  year      = {2016},
  doi       = {10.1126/science.aaf8834}
}

@article{Altman2021,
  title     = {Statistical mechanics of quantum error correction and the measurement-induced transition},
  author    = {Altman, Ehud and Vasseur, Romain and Potter, Andrew C.},
  journal   = {Phys. Rev. B},
  volume    = {104},
  number    = {13},
  pages     = {134204},
  year      = {2021},
  doi       = {10.1103/PhysRevB.104.134204}
}

@article{RevModPhys.78.275,
  title = {The kernel polynomial method},
  author = {Wei\ss{}e, Alexander and Wellein, Gerhard and Alvermann, Andreas and Fehske, Holger},
  journal = {Rev. Mod. Phys.},
  volume = {78},
  issue = {1},
  pages = {275--306},
  numpages = {0},
  year = {2006},
  month = {Mar},
  publisher = {American Physical Society},
  doi = {10.1103/RevModPhys.78.275},
  url = {https://link.aps.org/doi/10.1103/RevModPhys.78.275}
}

@article{noel2022measurement,
  title={Measurement-induced quantum phases realized in a trapped-ion quantum computer},
  author={Noel, Crystal and Niroula, Pradeep and Zhu, Daiwei and Risinger, Andrew and Egan, Laird and Biswas, Debopriyo and Cetina, Marko and Gorshkov, Alexey V and Gullans, Michael J and Huse, David A and others},
  journal={Nature Physics},
  volume={18},
  number={7},
  pages={760--764},
  year={2022},
  publisher={Nature Publishing Group UK London}
}

@article{koh2022experimental,
  title={Experimental realization of a measurement-induced entanglement phase transition on a superconducting quantum processor},
  author={Koh, Jin Ming and Sun, Shi-Ning and Motta, Mario and Minnich, Austin J},
  journal={arXiv preprint arXiv:2203.04338},
  year={2022}
}

@article{nandkishore2015many,
  title={Many-body localization and thermalization in quantum statistical mechanics},
  author={Nandkishore, Rahul and Huse, David A},
  journal={Annu. Rev. Condens. Matter Phys.},
  volume={6},
  number={1},
  pages={15--38},
  year={2015},
  publisher={Annual Reviews}
}

@article{imbrie2016many,
  title={On many-body localization for quantum spin chains},
  author={Imbrie, John Z},
  journal={Journal of Statistical Physics},
  volume={163},
  number={5},
  pages={998--1048},
  year={2016},
  publisher={Springer},
url={https://link.springer.com/article/10.1007/s10955-016-1508-x}
}

@article{li2018quantum,
  title={Quantum Zeno effect and the many-body entanglement transition},
  author={Li, Yaodong and Chen, Xiao and Fisher, Matthew PA},
  journal={Physical Review B},
  volume={98},
  number={20},
  pages={205136},
  year={2018},
  publisher={APS}
}

@article{Nahum2021PRXQ,
  title = {Measurement and Entanglement Phase Transitions in All-to-All Quantum Circuits, on Quantum Trees, and in Landau-Ginzburg Theory},
  author = {Nahum, Adam and Roy, Sthitadhi and Skinner, Brian and Ruhman, Jonathan},
  journal = {PRX Quantum},
  volume = {2},
  pages = {010352},
  year = {2021},
  doi = {10.1103/PRXQuantum.2.010352},
  url = {https://doi.org/10.1103/PRXQuantum.2.010352}
}

@article{Bao2020PRB,
  title = {Theory of the phase transition in random quantum circuits},
  author = {Bao, Yimu and Choi, Soonwon and Altman, Ehud},
  journal = {Physical Review B},
  volume = {101},
  pages = {104301},
  year = {2020},
  doi = {10.1103/PhysRevB.101.104301},
  url = {https://doi.org/10.1103/PhysRevB.101.104301}
}

@article{Gullans2020PRL,
  title = {Dynamical Purification Transition Induced by Measurements},
  author = {Gullans, Michael J. and Huse, David A.},
  journal = {Physical Review Letters},
  volume = {125},
  pages = {070606},
  year = {2020},
  doi = {10.1103/PhysRevLett.125.070606},
  url = {https://doi.org/10.1103/PhysRevLett.125.070606}
}

@article{Ippoliti2021PRX,
  title = {Entanglement Phase Transitions in Measurement-Only Dynamics},
  author = {Ippoliti, Matteo and Gullans, Michael J. and Gopalakrishnan, Sarang and Huse, David A. and Khemani, Vedika},
  journal = {Physical Review X},
  volume = {11},
  pages = {011030},
  year = {2021},
  doi = {10.1103/PhysRevX.11.011030},
  url = {https://doi.org/10.1103/PhysRevX.11.011030}
}

@article{Abanin2017RMP,
  title = {Effective Hamiltonians, prethermalization, and slow energy absorption in periodically driven many-body systems},
  author = {Abanin, D. A. and De Roeck, W. and Ho, W. W. and Huveneers, F. m. c.},
  journal = {Reviews of Modern Physics},
  volume = {89},
  pages = {011001},
  year = {2017},
  doi = {10.1103/RevModPhys.89.011001},
  url = {https://doi.org/10.1103/RevModPhys.89.011001}
}

@article{schreiber2015observation,
  title={Observation of many-body localization of interacting fermions in a quasirandom optical lattice},
  author={Schreiber, Michael and Hodgman, Sean S and Bordia, Pranjal and L{\"u}schen, Henrik P and Fischer, Mark H and Vosk, Ronen and Altman, Ehud and Schneider, Ulrich and Bloch, Immanuel},
  journal={Science},
  volume={349},
  number={6250},
  pages={842--845},
  year={2015},
  publisher={American Association for the Advancement of Science}
}

@article{bordia2017probing,
  title={Probing slow relaxation and many-body localization in two-dimensional quasiperiodic systems},
  author={Bordia, Pranjal and L{\"u}schen, Henrik and Scherg, Sebastian and Gopalakrishnan, Sarang and Knap, Michael and Schneider, Ulrich and Bloch, Immanuel},
  journal={Physical Review X},
  volume={7},
  number={4},
  pages={041047},
  year={2017},
  publisher={APS}
}

@article{kohlert2019observation,
  title={Observation of many-body localization in a one-dimensional system with a single-particle mobility edge},
  author={Kohlert, Thomas and Scherg, Sebastian and Li, Xiao and L{\"u}schen, Henrik P and Das Sarma, Sankar and Bloch, Immanuel and Aidelsburger, Monika},
  journal={Physical review letters},
  volume={122},
  number={17},
  pages={170403},
  year={2019},
  publisher={APS}
}

@article{zhu2021probing,
  title={Probing many-body localization on a noisy quantum computer},
  author={Zhu, Daiwei and Johri, S and Nguyen, NH and Alderete, C Huerta and Landsman, KA and Linke, NM and Monroe, C and Matsuura, AY},
  journal={Physical Review A},
  volume={103},
  number={3},
  pages={032606},
  year={2021},
  publisher={APS}
}

@article{roushan2017spectroscopic,
  title={Spectroscopic signatures of localization with interacting photons in superconducting qubits},
  author={Roushan, Pedram and Neill, Charles and Tangpanitanon, J and Bastidas, Victor M and Megrant, A and Barends, Rami and Chen, Yu and Chen, Z and Chiaro, B and Dunsworth, A and others},
  journal={Science},
  volume={358},
  number={6367},
  pages={1175--1179},
  year={2017},
  publisher={American Association for the Advancement of Science}
}

@article{chiaro2020growth,
  title={Growth and preservation of entanglement in a many-body localized system},
  author={Chiaro, Ben and Foxen, Brooks and McEwen, Matthew and Martinis, John},
  journal={Bulletin of the American Physical Society},
  volume={65},
  year={2020},
  publisher={APS}
}

@article{gong2021experimental,
  title={Experimental characterization of the quantum many-body localization transition},
  author={Gong, Ming and de Moraes Neto, Gentil D and Zha, Chen and Wu, Yulin and Rong, Hao and Ye, Yangsen and Li, Shaowei and Zhu, Qingling and Wang, Shiyu and Zhao, Youwei and others},
  journal={Physical Review Research},
  volume={3},
  number={3},
  pages={033043},
  year={2021},
  publisher={APS}
}

@article{kucsko2018critical,
  title={Critical thermalization of a disordered dipolar spin system in diamond},
  author={Kucsko, Georg and Choi, Soonwon and Choi, Joonhee and Maurer, Peter C and Zhou, Hengyun and Landig, Renate and Sumiya, Hitoshi and Onoda, Shinobu and Isoya, Junich and Jelezko, Fedor and others},
  journal={Physical review letters},
  volume={121},
  number={2},
  pages={023601},
  year={2018},
  publisher={APS}
}

@article{agrawal2024observing,
  title={Observing quantum measurement collapse as a learnability phase transition},
  author={Agrawal, Utkarsh and Lopez-Piqueres, Javier and Vasseur, Romain and Gopalakrishnan, Sarang and Potter, Andrew C},
  journal={Physical Review X},
  volume={14},
  number={4},
  pages={041012},
  year={2024},
  publisher={APS}
}

@article{feng2025postselection,
  title={Postselection-free experimental observation of the measurement-induced phase transition in circuits with universal gates},
  author={Feng, Xiaozhou and C{\^o}t{\'e}, Jeremy and Kourtis, Stefanos and Skinner, Brian},
  journal={arXiv preprint arXiv:2502.01735},
  year={2025}
}

@article{google2023measurement,
  title={Measurement-induced entanglement and teleportation on a noisy quantum processor},
  journal={Nature},
  volume={622},
  number={7983},
  pages={481--486},
  year={2023},
  publisher={Nature Publishing Group UK London}
}

\section*{End Matter}

\subsection*{Unitary time evolution}
\begin{figure}[H]
    \centering
    \includegraphics[width=\linewidth]{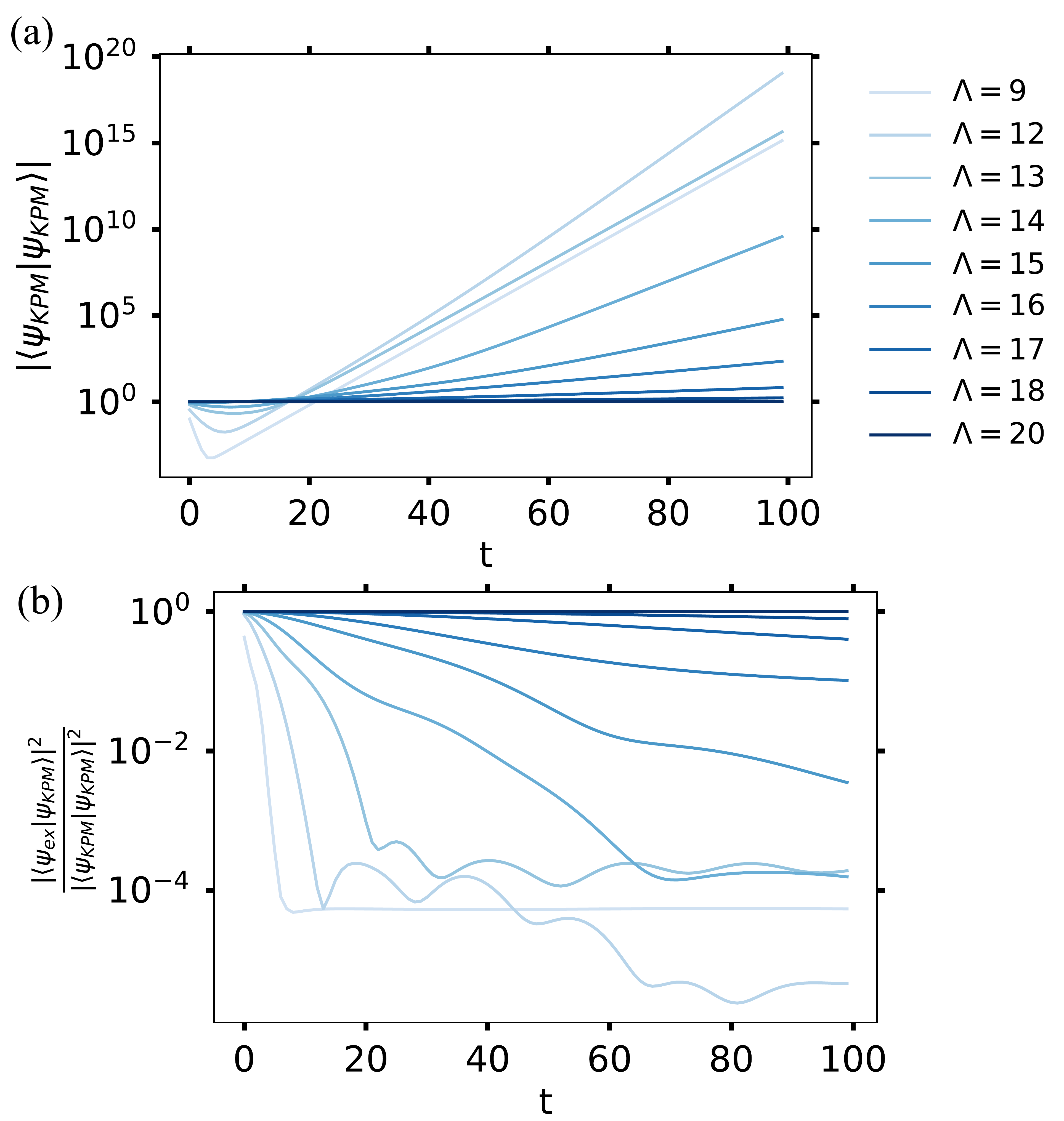}
    \caption{We evolve the state $\ket{\psi_0}=\otimes_{i=1}^L \ket{\uparrow}$ with the Hamiltonian~\eqref{Hmbl} for system size $L=8$ and disorder strength $W=5$. (a) Norm of  $\ket{\psi_\text{KPM}}$  as a function of time for different truncation orders $\Lambda$. (b) Fidelity of the KPM wavefunction $\ket{\psi_\text{KPM}}$ for different $\Lambda$ as compared with the exact time evolved  state $\ket{\psi_\text{ex}}$.   }
    \label{fig:KPM}
\end{figure}
Instead of diagonalizing the microscopic Hamiltonian, which becomes intractable for large systems, we approximate the exponential $U=e^{-iH\delta t}$ using the kernel polynomial method (KPM) following Ref.~\cite{fehske2007computational}. 
This method utilizes Chebyshev polynomials $T_n(x)$ defined on the interval $x \in [-1,1]$. 
Therefore, we first rescale the Hamiltonian $H$ so that its spectrum lies within this interval:
\begin{equation}
\tilde{H} = \frac{H - b}{a}, 
\quad a = \frac{E_{\text{max}} - E_{\text{min}}}{2}, 
\quad b = \frac{E_{\text{max}} + E_{\text{min}}}{2},
\end{equation}
where $E_{\text{min}}$ and $E_{\text{max}}$ denote the minimum and maximum eigenvalues of $H$.

With this rescaling, the time evolution operator can be written as
\begin{equation}
U(t) = e^{-iHt} = e^{-ibt}\, e^{-iat\tilde{H}}.
\end{equation}
The global phase factor $e^{-ibt}$ is exact, while the nontrivial part $e^{-iat\tilde{H}}$ is expanded in terms of Chebyshev polynomials to an order $\Lambda$:
\begin{equation}
e^{-iat\tilde{H}} \approx J_0(at)I \;+\; 2\sum_{n=1}^{\Lambda} (-i)^n J_n(at)\, T_n(\tilde{H}),
\end{equation}
where $J_n$ are Bessel functions of the first kind.  

 Chebyshev polynomials are generated recursively according to
\begin{equation}
T_0(x) = 1, \quad 
T_1(x) = x, \quad 
T_{n+1}(x) = 2xT_n(x) - T_{n-1}(x).
\end{equation}
Thus, for any initial state $\ket{\psi}$, the propagated state at time $t$ can be approximated as
\begin{equation}
\ket{\psi(t)} \approx e^{-ibt}\biggl[J_0(at)\ket{\phi_0} 
+ 2\sum_{n=1}^{\Lambda} (-i)^n J_n(at)\ket{\phi_n}\biggr],
\end{equation}
with the vectors $\{\ket{\phi_n}\}$ defined recursively by
\begin{equation}
\ket{\phi_0} = \ket{\psi}, 
\quad \ket{\phi_1} = \tilde{H}\ket{\psi}, 
\quad \ket{\phi_{n+1}} = 2\tilde{H}\ket{\phi_n} - \ket{\phi_{n-1}}.
\end{equation}
This expansion reduces the problem to a sequence of sparse matrix–vector multiplications, 
avoiding the direct exponentiation of a large $2^N \times 2^N$ matrix.

We gauge the accuracy of the Chebyshev expansion through the  fidelity of the KPM. We compare the time-evolved state obtained 
from exact propagation, 
$|\psi_{\mathrm{ex}}(t)\rangle =  e^{-iHt}\,|\psi_0\rangle$
with the one obtained from the Chebyshev expansion, 
$|\psi_{\mathrm{KPM}}(t)\rangle = U_{\mathrm{KPM}}\,|\psi_{\mathrm{KPM}}(t-1)\rangle$ by
 evaluating the overlap 
$F(t) = \langle \psi_{\mathrm{ex}}(t) | \psi_{\mathrm{KPM}}(t) \rangle$.
 The two methods agree when $\Lambda\gtrsim 20$ as shown in Fig.~\ref{fig:KPM}. In the calculations presented in the main text, we fix $\Lambda = 400$, for which the results are indistinguishable from those obtained by exact time evolution.

\subsection*{Level spacing ratio of the microscopic model}\label{Level-spacing}
\begin{figure}[H]
    \centering
    \includegraphics[width=0.9\linewidth]{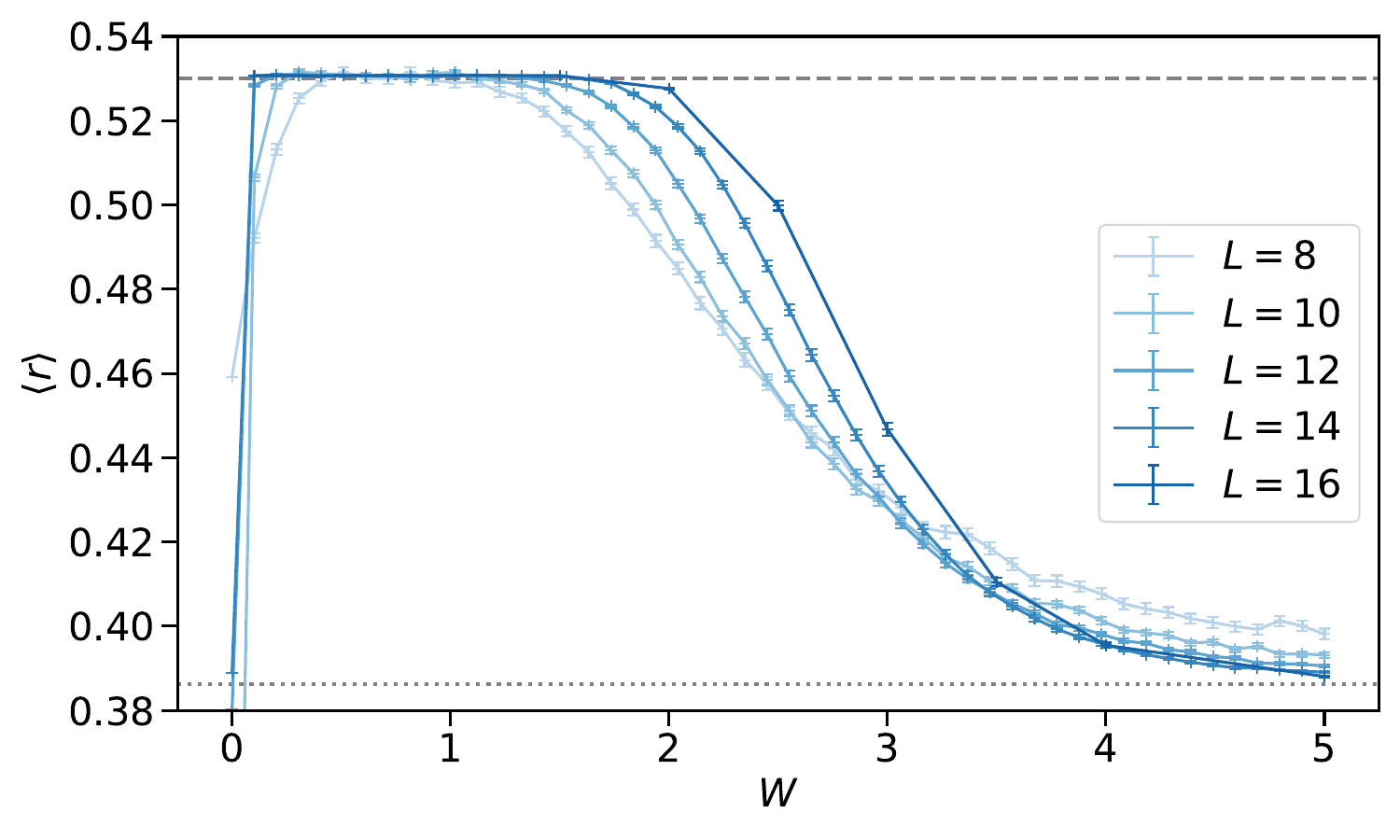}
    \caption{Average level-spacing ratio $\langle r \rangle$ for the Hamiltonian~\eqref{Hmbl} as a function of disorder strength $W$ for several
system sizes $L$. 
The dashed gray line indicates the Gaussian orthogonal ensemble value 
$\langle r \rangle \approx 0.53$, while the dotted gray line marks the 
Poisson value $\langle r \rangle = 2 \ln 2 - 1$.}
    \label{fig:level-spacing}
\end{figure}

To diagnose the many-body localization transition, we study the spectral statistics of the microscopic Hamiltonian~\eqref{Hmbl}. First, we obtain the ordered eigenvalues 
\begin{equation}
    E_0 < E_1 < E_2 < \cdots < E_{2^L - 1},
\end{equation}
by exact diagonalization.  We define the consecutive level spacings
\begin{equation}
    \delta_i = E_{i+1} - E_i,
\end{equation}
and the level spacing ratios
\begin{equation}
    r_i = \frac{\min(\delta_i, \delta_{i-1})}{\max(\delta_i, \delta_{i-1})}, \quad i=1,2,\dots,2^L-2.
    \label{eq:rdef}
\end{equation}

The average value $\langle r \rangle$ over disorder realizations (500 samples) and a chosen energy window (middle $60\%$) serves as a sensitive probe of localization in a finite size system.  For $W=0$, the model is integrable and hence level spacings follow  Poisson distribution with the average ratio $\braket{r}\sim 0.39$, see Fig.~\ref{fig:level-spacing}. As $W$ increases, the system becomes chaotic and hence level spacings conform to the Gaussian orthogonal ensemble (GOE) with average ratio $\braket{r}\sim 0.53$. As $W$ increases further, the level spacing distribution reverts back to  Poisson, indicating that the system is now in an MBL  phase. From the crossing of the curves for various $L$, we infer that a transition to the localized phase occurs at the critical value of $W_c\gtrsim 3.2$.

Note, however, that while this computation performed for a finite system seems to indicate a transition at a small value of $W_c=3.2$, more careful studies have revealed that this is in fact an extended prethermal crossover regime, and true many-body localization, if it occurs, will take place at a much larger disorder strength.

\subsection*{Critical measurement rate}
\begin{figure}
\includegraphics[width=0.8\linewidth]{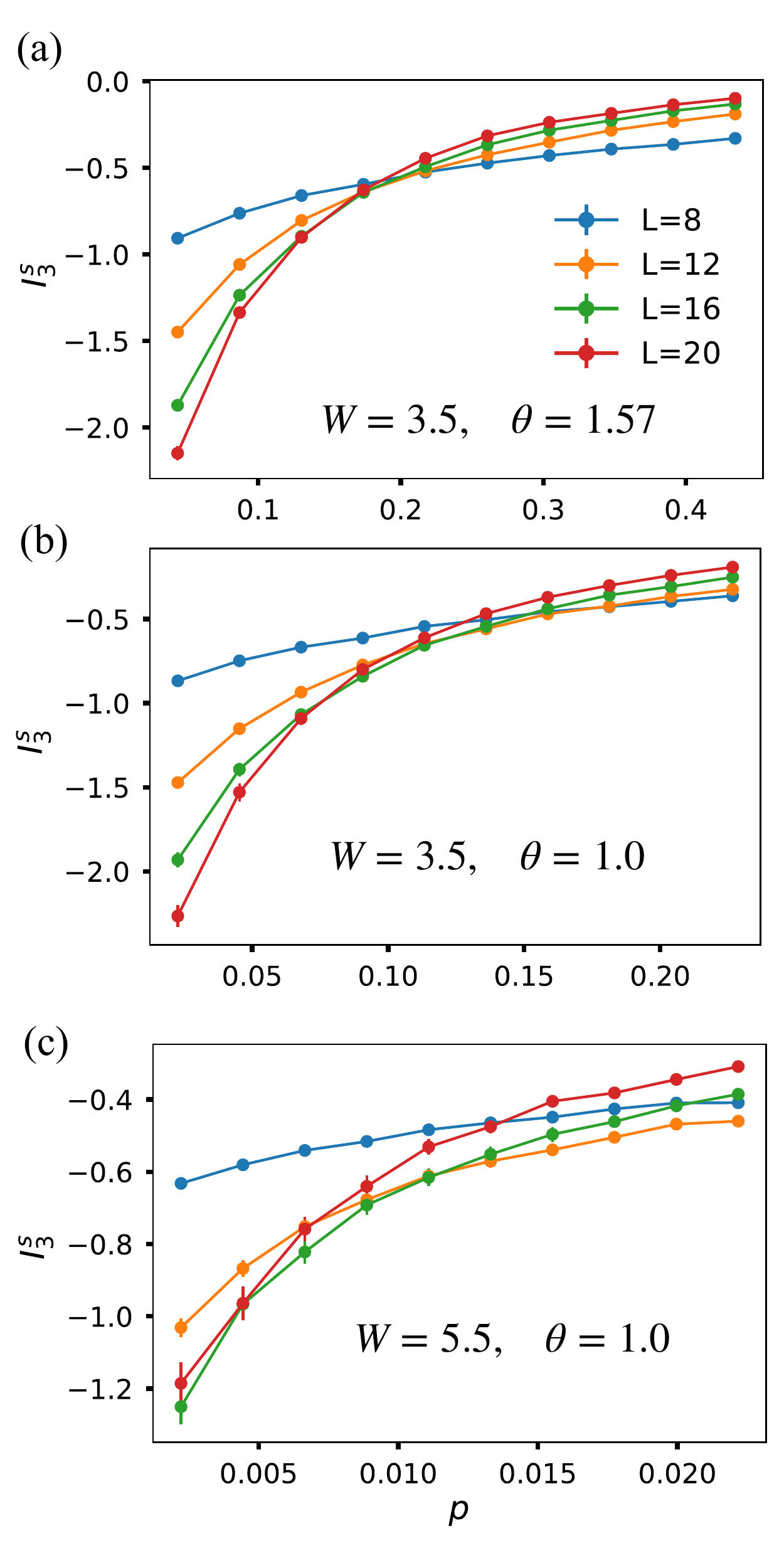}
    \caption{Three examples of crossings of plots of the tripartite mutual information $I_3^s$ vs. the measurement rate $p$. Observe  that  finite-size effects become stronger as the disorder strength $W$ increases or the measurement angle $\theta$ decreases. (a) $W=3.5$ and $\theta = 1.57$.  Finite size effects are negligible. (b) $W=3.5$ and $\theta = 1.0$. The $L=8$ plot is not crossing  at the same point as the plots for other $L$. (c) $W=5.5$ and $\theta = 1.0$. Finite size effects  are strongest and there is no clearly discernible crossing point. }
    \label{fig:expc}
\end{figure}
To evaluate the critical measurement rate $p_c$ for the microscopic model, we calculate the averaged tripartite mutual information $I_3$ as discussed in the main text. In Fig.~\ref{fig:expc}, we show three representative examples of the data crossings for $I_3$. As $\theta\to0$ and $W\to\infty$, the critical measurement rate  $p_c\to 0$ and the   finite-size effects become stronger. As   the finite size effects are very large for $L=8$, we extrapolate the critical measurement rate from $L=12, 16,$ and $20$ using the 
best fit to the scaling function $I_3(t\rightarrow \infty)\sim f\left((p-p_c)L^{1/\nu}\right)$. In cases  such as   Fig.~\ref{fig:expc}(c), when the crossing is difficult to discern, we marked the corresponding $p_c$ with a hollow dot in 
Fig.~\ref{fig:p_c}.

\begin{figure}
    \centering
    \includegraphics[width=0.9\linewidth]{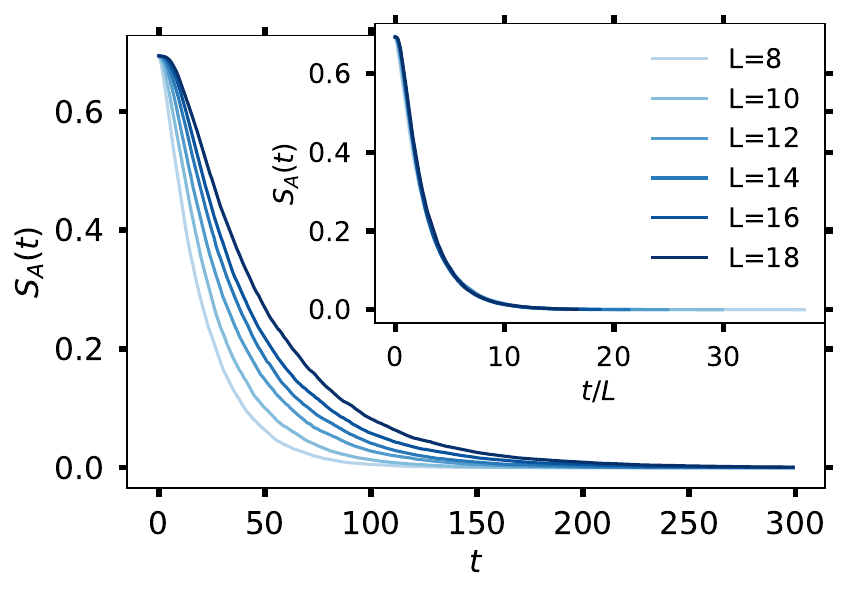}
    \caption{Average ancilla entropy $S_A(t)$ for the $l$-bit model at the critical measurement rate $p=p_c$. The inset shows that  plots  for different system sizes $L$ collapse onto a single curve $S_A(t/L)$ indicating that the dynamical exponent $z=1$.}
    \label{fig:Ancilla}
\end{figure}

\subsection*{Ancilla}

To determine the dynamical exponent $z$ at criticality, $p=p_c$, we calculate the average ancilla entropy $S_A(t)=\bigl[\bar{S}^{\{\eta,m_i(t)\}}_A(t)\bigr]$ for the $l$-bit model coupled to an ancilla $A$.  Fig.~\ref{fig:Ancilla} shows that the ancilla entropy obeys the scaling $S_A(t,L)=S_A(t/L)$, where $L$ is the system size. We therefore conclude that $z=1$.

To calculate $S_A$ numerically, we start from an initial state $\frac{1}{\sqrt{2}}(|0\rangle_A \otimes |\phi_1\rangle +|1\rangle_A \otimes |\phi_2\rangle)$, with $\langle \phi_1|\phi_2\rangle=0$. Here $A$ stands for the ancilla and $\ket{\phi}$ is the state of the system. The system is evolved under the protocol discussed in the main text. At each time step, we calculate the entanglement entropy between the ancilla and the system $S^{\{\eta,m_i(t)\}}_A(t)=-\rho_A\ln\rho_A$.  In Fig.~\ref{fig:Ancilla}, we  show the ancilla entropy $S_A(t)=\bigl[\bar{S}^{\{\eta,m_i(t)\}}_A(t)\bigr]$ averaged over 2000 quantum trajectories and realizations of the random variables $\eta$.

\end{document}